\documentclass[printer, traditabstract]{aa}
\usepackage{setspace}
\usepackage{txfonts} 
\usepackage{textcomp}
\usepackage{natbib}
\usepackage{amssymb}
\usepackage{graphicx}
\usepackage{tabularx}
\usepackage[T1]{fontenc}
\usepackage{multirow}
\usepackage{multirow}

\def\HI{H{\,\small I}}
\newcommand{\sauron}{{\texttt {SAURON}}}

\def\HI{H{\,\small I}}

\def\HI{H{\,\small I}}

\def\emph#1{{\sl #1}}
\newcommand{\ltsima} {$\; \buildrel < \over \sim \;$}
\newcommand{\gtsima} {$\; \buildrel > \over \sim \;$}
\newcommand{\lta} {\lower.5ex\hbox{\ltsima}}
\newcommand{\gta} {\lower.5ex\hbox{\gtsima}}

\newcommand{\atlas}{{ATLAS$^{\rm 3D}$}}

\begin{document}

\title{From star-forming galaxies to AGN: \\
the global HI content from a stacking experiment}

\author{K. Ger\'{e}b $^{1,2,3}$, R. Morganti$^{1,2}$, T.A. Oosterloo$^{1,2}$, L. Hoppmann$^{4}$, L. Staveley-Smith$^{4,5}$
}

\authorrunning{Ger\'{e}b et al.}

\titlerunning{The global \HI\ content of SF galaxies and AGN}

\institute{$^{1}$ASTRON, the Netherlands Institute for Radio Astronomy, Postbus 2, 7990 AA, Dwingeloo, The Netherlands\\ 
$^{2}$Kapteyn Astronomical Institute, University of Groningen, P.O. Box 800, 9700 AV Groningen, The Netherlands \\
$^{3}$Centre for Astrophysics $\&$ Supercomputing, Swinburne University of Technology, Hawthorn, VIC 3122, Australia \\
$^{4}$International Centre for Radio Astronomy Research (ICRAR), M468, University of Western Australia, 35 Stirling Hwy, WA 6009, Australia\\
$^{5}$ARC Centre of Excellence for All-sky Astrophysics (CAASTRO), Australia }


\abstract{We study the atomic neutral hydrogen (\HI) content of $\sim$1600 galaxies up to $z \sim 0.1$ using stacking techniques. The observations were carried out with the Westerbork Synthesis Radio Telescope (WSRT) in the area of the SDSS South Galactic Cap (SSGC), where we selected a galaxy sample from the SDSS spectroscopic catalog. Multi-wavelength information is provided by SDSS, NVSS, GALEX, and WISE. We use the collected information to study \HI\ trends with color, star-forming, and active galactic nuclei (AGN) properties.

Using NUV $-$ $r$ colors, galaxies are divided into blue cloud, green valley and red sequence galaxies. As expected based on previous observations, we detect \HI\ in green valley objects with lower amounts of \HI\ than blue galaxies, while stacking only produces a 3-$\sigma$ upper limit for red galaxies with M$_{\rm HI}$ $<$ (5 $\pm$ 1.5) $\times$ 10$^{8}$ M$_{\odot}$ and M$_{\rm HI}/\rm{L}_r$  $<$ 0.02 $\pm$ 0.006 $\rm M_{\odot}  / \rm L_{\odot} $ (averaged over four redshift bins up to $z \sim 0.1$). 
We find that the \HI\ content is more dependent on NUV $-$ $r$ color, and less on ionization properties, in the sense that regardless of the presence of an optical AGN (based on optical ionization line diagnostics), green-valley galaxies always show \HI, whereas red galaxies only produce an upper limit. This suggests that feedback from optical AGN is not the (main) reason for depleting large-scale gas reservoirs.

Low-level radio continuum emission in our galaxies can stem either from star formation, or from AGN. We use the WISE color-color plot to separate these phenomena by dividing the sample into IR late-type and IR early-type galaxies. We find that the radio emission in IR late-type galaxies stems from enhanced star formation, and this group is detected in \HI. However, IR early-type galaxies lack any sign of \HI\ gas and star formation activity, suggesting that radio AGN are likely to be the source of radio emission in this population.

Future \HI\ surveys will allow for extending our studies to higher redshift, and for testing any possible evolution of the \HI\ content in relation to star-forming and AGN properties up to cosmologically significant distances. Such surveys will provide enough data to test the effect of radio/optical AGN feedback on the \HI\ content at lower, currently rather unexplored \HI\ detection limit (M$_{\rm HI}$ < 10$^7$ M$_{\odot}$).  }

\maketitle

\section{Introduction}\label{Intro}

The amount and conditions of cold \HI\ gas in galaxies are, in a direct or indirect way, related to star formation (SF) processes and to black hole fuelling. Therefore, our knowledge of the \HI\ properties is crucial for our understanding of the intricate processes of galaxy formation and evolution.
Our knowledge of the gas content in various types of galaxies in the nearby Universe has increased substantially thanks to large single-dish \HI\ surveys such as the \HI\ Parkes All Sky Survey (HIPASS) \citep{Meyer2004, Zwaan2005}, the Arecibo Legacy Fast ALFA (ALFALFA) survey \citep{Giovanelli2005, Haynes2011} and detailed imaging surveys like WHISP \citep{Hulst}, THINGS \citep{THINGS}, \sauron\ \citep{Morganti2006, Oosterloo2010} and \atlas\ \citep{Serra}.

Scaling relations linking the \HI-to-stellar mass ratio to global physical properties of galaxies, e.g. stellar mass, stellar mass surface density and colour, have been derived for massive galaxies in the Galex Arecibo SDSS Survey (GASS, \citealt{Catinella2010, Cortese2011}). 
These studies show that the \HI\ gas fraction most strongly correlates with NUV $-$ $r$ color and stellar surface mass density, $\mu_{\star}$, which relation is most naturally explained as being the Kennicutt-Schmidt law \citep{Kennicutt1998, Zhang2009}. The correlation is not surprising, since the NUV $-$ $r$ color and $\mu_{\star}$ correlate with specific star formation rate and morphological type respectively, which properties are expected to be closely tied to gas fraction.   
Blue galaxies, in general, are \HI\ rich and actively forming stars, whereas red galaxies tend to be more \HI-poor or depleted of gas. The `green valley' is considered to be a transition population between star-forming blue and quiescent red galaxies \citep{Martin2007, Wyder, Schimi2007}, displaying residual star-formation signatures \citep{Yi}. As a results, green, intermediate colors have been interpreted as a sign for the recent quenching of star formation in this population \citep{Salim2007}.

The mechanisms that drive quenching processes in galaxies are yet to be understood. 
It is thought that feedback processes are important in affecting the gas reservoirs and consequently the star formation processes, particularly in massive, bulgy galaxies \citep{Croton2006, deLucia2007, Somerville2008}. Theoretical studies emphasize the role of AGN in quenching star formation processes by heating the gas reservoirs \citep{Croton2006, deLucia2007, Somerville2008, Booth2009Feedb, Debuhr2012}, or by driving the gas outside of the host galaxy \citep{diMatteo2005, Hopkins2006Feedb, Wagner}. Therefore, observational evidence is of key importance for our understanding of the effects that AGN can have on the (neutral) gas and star formation processes in galaxies.

Active galactic nuclei (AGN) can reveal their presence in many ways, e.g., optical emission lines in the nuclear region, and radio jet activity. AGN are often selected optically, based on Baldwin, Phillips \& Terlevich (BPT) [\ion{O}{III}]/H$\beta$ vs. [\ion{N}{II}]/H$\alpha$ line ratio diagnostics \citep{Baldwin}.
As \cite{Best2005} pointed out, optical line selection naturally leads to a bias in favour of AGN with a rich inter-stellar medium. This happens for the reason that to optically identify an AGN, the presence of black holes and sufficient amounts of gas that produce the bright emission are both necessary. However, AGN which are selected in the radio continuum often show only weak or no emission lines, and this population would be missed by optical selection. 

Radio emission in galaxies can either stem from AGN activity, or it can also originate from star formation processes. Therefore, care has to be taken to separate these phenomena when looking for AGN in the radio continuum. From the 1.4 GHz radio luminosity function it is expected that radio AGN become dominant over star formation at radio power \mbox{log(P / W Hz$^{-1}$) $>$ 23} (Mauch and Sadler 2007). As a result, the highest radio power sources are most likely associated with active galactic nuclei (AGN). However, the nature of the faint radio population is controversial. Low radio luminosity sources have been identified as being either SF galaxies (starbursts, spirals or irregulars), or low power radio-loud and/or radio-quiet AGN (e.g. faint FR I, Seyfert galaxies). 
Many different approaches have been taken to identify SF and AGN processes in low luminosity radio sources (see Prandoni et al. 2009 for an overview). 
Among these, IR color-color diagrams were found to be efficient in disentangling these two phenomena.

In order to learn about the SF and AGN processes of different galaxy populations in relation to their \HI\ gas content, we need large samples to globally look at the \HI\ properties of galaxies. Large samples can be achieved either by large sky surveys such as ALFALFA, HIPASS, or by looking at smaller portions of the sky while extending the observations to higher redshift.
The limitation of the second approach is that at higher redshift, \HI\ studies are limited by sensitivity and bandwith, and such observations are quite expensive in terms of observation time even with the most sensitive telescopes \citep{Catinella2015}.  However, alternative techniques, like spectral stacking, have been efficiently used to measure the average \HI\ content of galaxies in the relatively distant Universe. 
In combination with multiwavelength data, stacking has also provided a powerful tool to study the neutral gas properties in different galaxy populations over the last years \citep{Lah2007,Lah2009, Fabelloa, Fabellob, Verheijen, Delhaize, Gereb2013}.

In a previous paper (\citealt{Gereb2013}, i.e. Paper 1.) we carried out spectral stacking on galaxies selected in the Lockman Hole (LH) field. We reported the detection of \HI\ gas not only in normal SF galaxies, but also in low ionization nuclear emission region (LINER) galaxies \citep{Kauffmann, Kewley2006, Best2012}. We show, in good agreement with \sauron\ and \atlas, that albeit in lower amounts, \HI\ and star formation are present not just in typical SF galaxies (generally blue, late type spirals), but also in galaxies with older stellar populations (redder), or in galaxies hosting AGN.
In Paper 1. we successfully separated red ($g - r >$ 0.7) star-forming  galaxies from potential low-luminosity radio AGN using IR diagnostics. We found that the two groups are well separated in terms of \HI\ properties as well. The main limitation of the LH study was that our selection of different sub-samples was restricted by the small sample size.

In this paper we present the \HI\ properties of a larger sample, $\sim$ 1600 galaxies. We confirm several trends derived by \cite{Gereb2013} and expand on these results. This work is made possible by the increase, by more than an order of magnitude, in the number of galaxies, allowing us to lower the detection limit by a factor of 4 using stacking techniques. Furthermore, the relatively high spatial resolution (see Section \ref{Sec:Obs}) of the observations used in this study reduces the risk of confusion from companion galaxies, often present in previous, single-dish stacking experiments. The larger sample makes it possible to study in more detail the \HI\ content of different galaxy populations. 
In addition to \cite{Gereb2013}, we study galaxies where quenching and feedback are thought to be affecting the gas reservoirs, e.g. green valley objects, AGN (selected based on optical emission lines).

A sub-sample of our galaxies is associated with radio emission in the NRAO VLA Sky Survey (NVSS, \citealt{Condon1998}). 
We explore the SF and AGN properties of the radio sources in relation to their \HI\ content, and we discuss the possibility of separating these phenomena using IR colors.

In a second paper, using the same dataset, the $\Omega_{\rm{HI}}$ will be investigated in the same redshift range (Hoppmann et al., in prep.).

In this paper the standard cosmological model is used, with parameters $\Omega_{m}$ = 0.3, $\Lambda$ = 0.7 and $H_0$ = 70 km s$^{-1}$ Mpc$^{-1}$.

\begin{figure*}
    \begin{center}
       \includegraphics[width=0.43\textwidth]{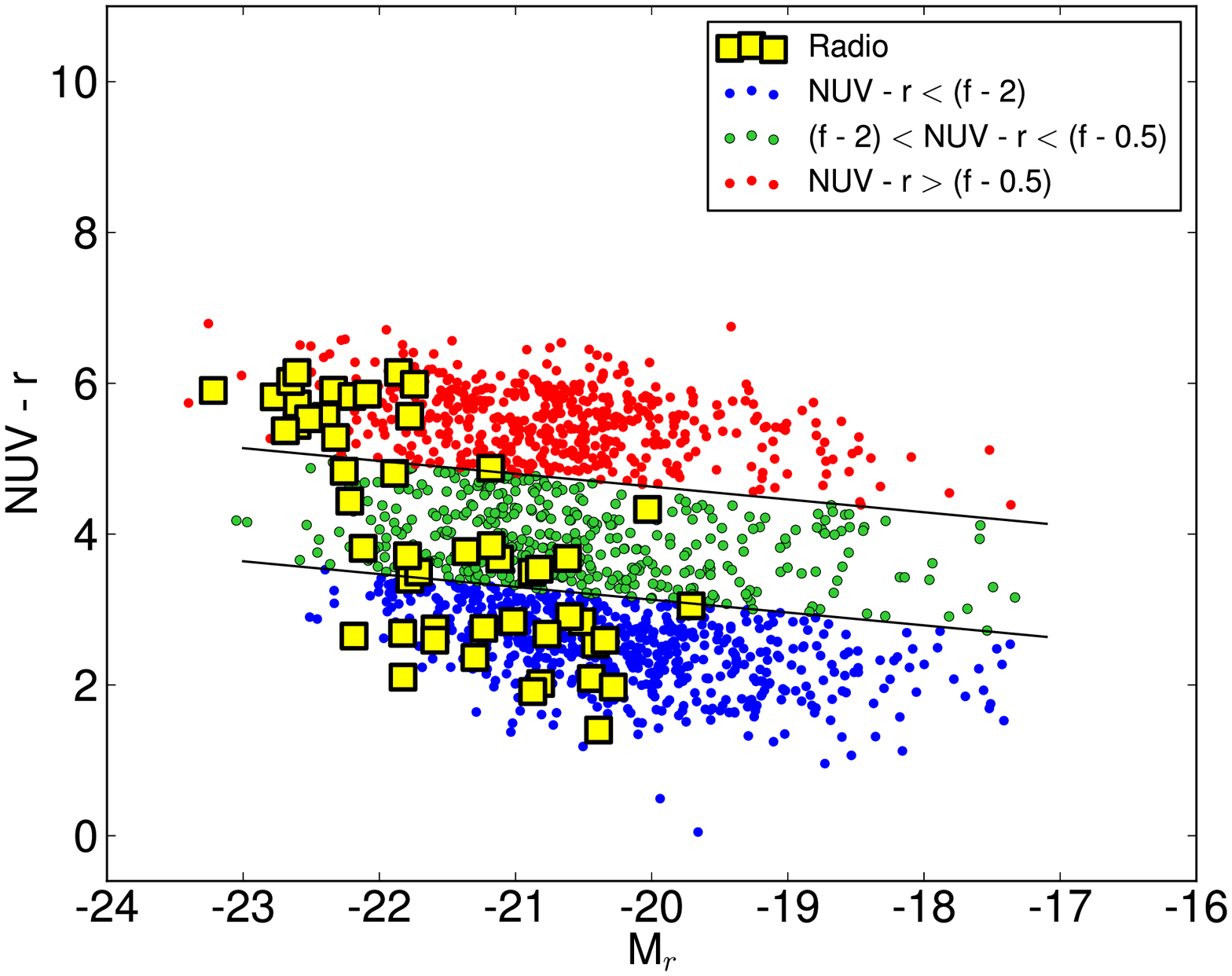}
       \includegraphics[width=0.43\textwidth]{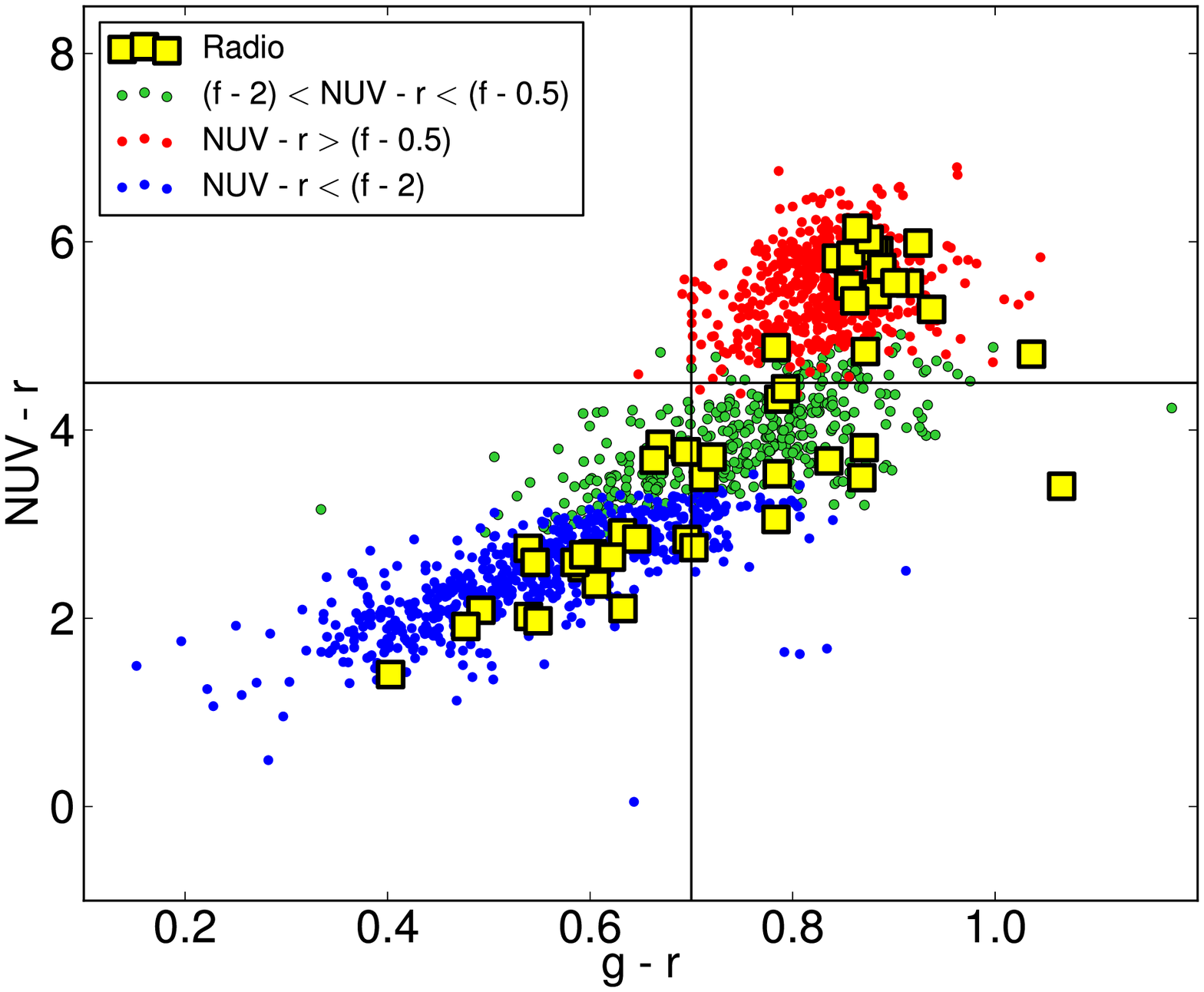}
   \end{center}
\caption{ 1. Color-magnitude diagram (left) 2. UV-optical color-color plot (right); The galaxies are color coded according to the color-magnitude selection by \cite{Wyder}, i.e. blue cloud, green valley and the red sequence. NVSS radio sources are marked by yellow squares.}\label{fig:CMD}
\end{figure*}

\begin{figure*}
    \begin{center}
       \includegraphics[width=0.46\textwidth]{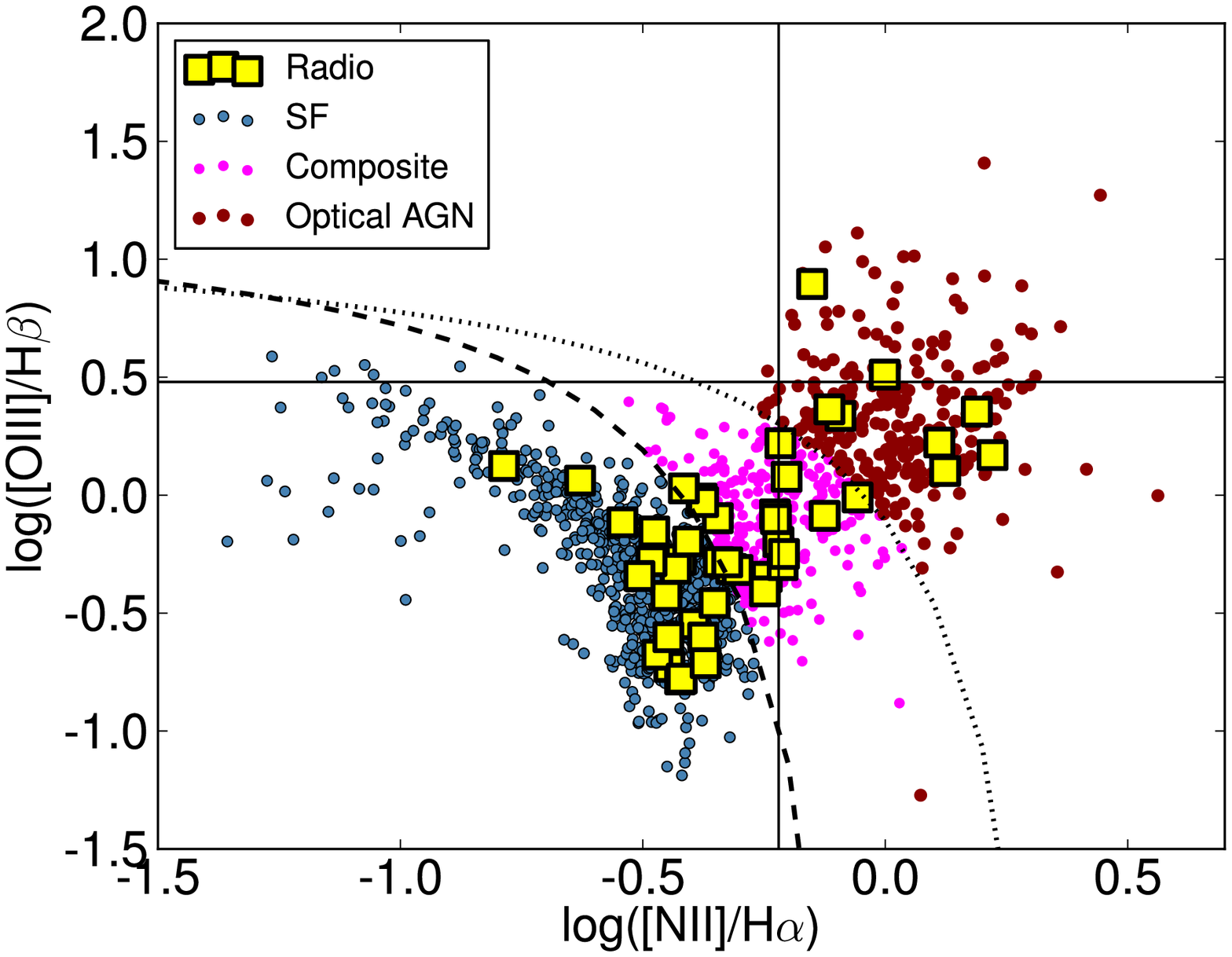}
       \includegraphics[width=0.43\textwidth]{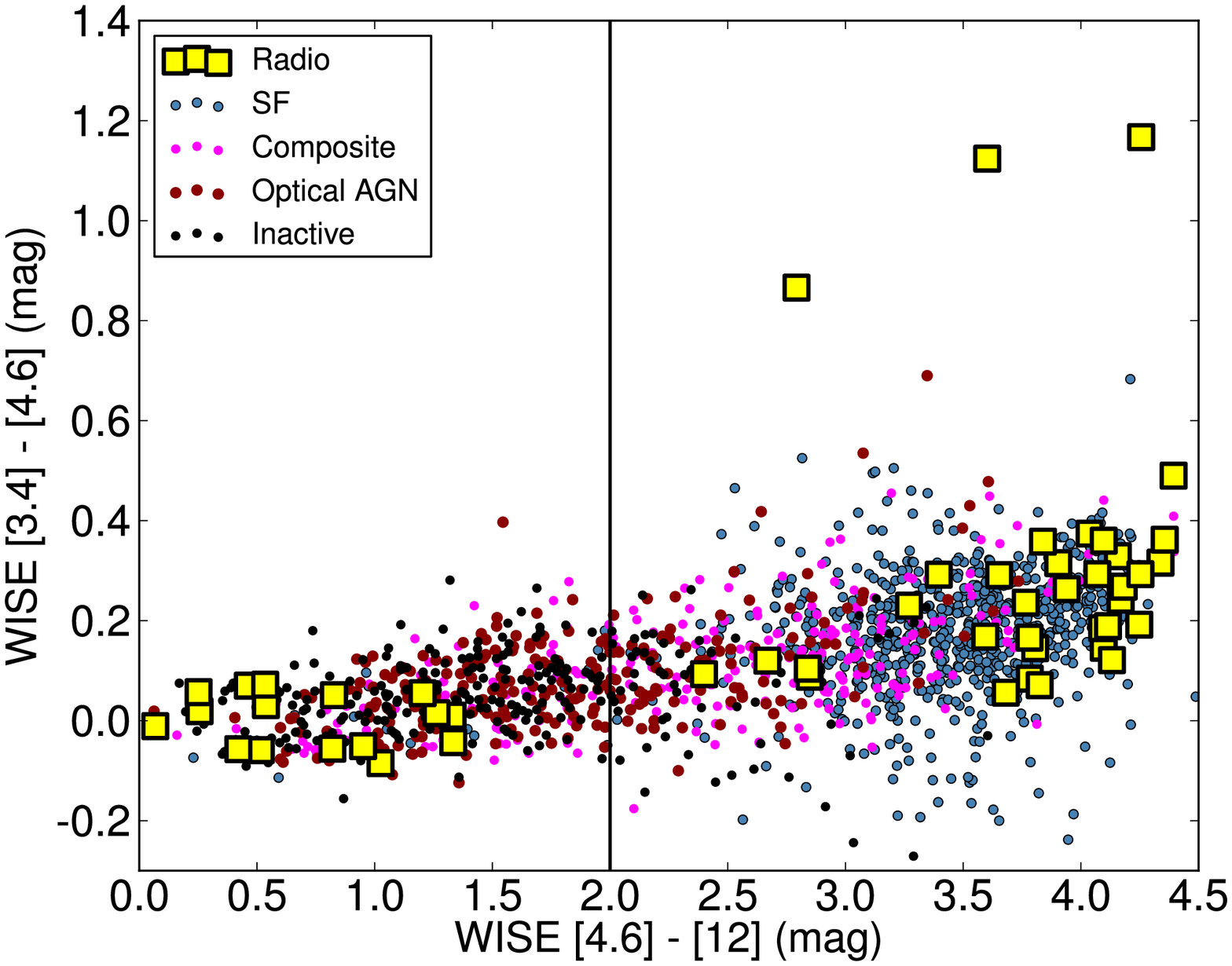}
   \end{center}
\caption{1. BPT diagram (left). The dashed line \citep{Kauffmann} is separating SF galaxies (below the dashed line) from composite galaxies (between dashed and dotted line). Optical AGN are located above the dotted line \citep{Kewley, Kewley2006}. Inactive galaxies do not appear in the diagram. 
2. WISE IR color-color plot (right). The sources are color-coded according to the BPT selection. The vertical solid line is separating IR early-type ([4.6$\mu m$] - [12$\mu m$]  < 2) and IR late-type galaxies ([4.6$\mu m$] - [12$\mu m$]  > 2). 
Radio sources are marked by yellow squares.}\label{fig:BPT}
\end{figure*}

\section{Observations and sample selection}\label{Sec:Obs}

The \HI\ observations were carried out with the Westerbork Synthesis Radio Telescope (WSRT) at 1.4 GHz, in the period May 2011 - October 2012, in the area of the SDSS South Galactic Cap (SSGC). Between the coordinates 21$^{h}$ $<$ RA $<$ 2$^{h}$, 10$^{\circ}$ $<$ DEC $<$ 16$^{\circ}$ (J2000), 35 WSRT pointings were observed and used for \HI\ spectral stacking. The SSGC was chosen because in this sky area the oversubscription of the WSRT is usually the lowest.

The redshift range 0 $\textless$ $z$ $\textless$ 0.12 is covered by $8 \times 20$ MHz bands with 128 frequency channels in each band (1280 - 1420 MHz, the bands overlap 3 MHz). The corresponding velocity resolution is 38 km s$^{-1}$. The integration time is 12 hours for most of the observations. The synthesized beam is typically 70 $\times$ 9 arcsec. The elongated beam shape is (mainly) the result of the low declination observations with the East-West WSRT array. However, the wealth of ancillary data available for this region of the sky outweighed the disadvantages of having such an elongated beam.

We use the Sloan Digital Sky Survey (SDSS, York et al. 2000) to define the central coordinates of the 35 WSRT pointings. These were selected in the sightline of high density SDSS regions with the goal of maximizing the number of detections. Because of the relatively large field of view and redshift range, both low and high density environments are well sampled.
The observed spectroscopic sample is cross-correlated with the Galaxy Evolution Explorer (GALEX, Martin et al. 2005), providing 1595 galaxies that can be used for stacking. We also use the Wide-Field Infrared Survey Explorer (WISE, \citealt{Wright}) to obtain infrared (IR) data for our galaxies. The WISE sample is complete to the 99 percent level. In addition, a few (50, within a search radius of 15 arcsec) galaxies are identified with radio counterparts in the NVSS  survey. The collected multiwavelength data allows us to combine various galaxy parameters and investigate the \HI\ properties of different galaxy populations.

\section{Data reduction and \HI\ stacking}\label{stacking}

The data were reduced using the MIRIAD package \citep{Sault}. Bad data were flagged from the datasets, with extra care for the most prominent RFI in the  lowest-frequency band (1280 - 1300 MHz).

The standard way to subtract the continuum is by fitting a low-order polynomial to the line-free channels. Our datasets cover a broad, 140 MHz bandwidth composed by 8 spectral windows. We fit each spectral band separately; however, in cases where the \HI\ line is at the edge of the band, instead of interpolating we need to extrapolate due to lack of continuum at the edge. This makes the  polynomial fitting unsuccessful and the continuum subtraction more uncertain. These \HI\ lines create a dip in the stacked spectra because of the bad continuum subtraction. 

To avoid this effect, first we perform a continuum subtraction in the {\it uv}-plane, using the clean components of each field. The clean components were created as a result of deconvolution of the continuum images with the dirty beam. This subtraction step removes most of the continuum from the \HI\ data cubes. However, we perform a second continuum subtraction by fitting a second-order polynomial to the spectra to subtract low-level residual continuum emission coming from very bright sources. Following these steps, we eliminate the dip from the stacked spectra.

Stacking is done similarly as described in \cite{Gereb2013}, centred on the redshift of the galaxy to be stacked. Galaxies are stacked in four redshift ranges, between 0.02 $<$ $z$ $<$ 0.12 each redshift range covers $\Delta z$ = 0.02 (except the highest redshift range, 0.08 $<$ $z$ $<$ 0.12). After noise corrections in every channel, the weighted sum of the stacked spectra is:

\begin{equation}\label{eq:weighting}
 S( \nu ) = \frac{ \sum \frac{ S_{i}(\nu) } {\sigma_{i}^{2} (\nu)} }  {\sum (1 / \sigma_{i}^{2} (\nu) ) } 
\end{equation}

\noindent
where $S (\nu)$ is the stacked spectrum, $\sigma_{i}(\nu)$ and $S_{i}(\nu)$ are the channel noise and the extracted spectrum of source $i$ as a function of frequency. \\

\noindent
The \HI\ masses are derived from the formula: 

\begin{equation}
\frac{M_{\rm HI}} {M_{\odot}} = \frac{2.356 \times 10^{5}}{(1+z)}  \bigg (\frac{S_{\nu}} {\rm{Jy}}\bigg)    \bigg(\frac{D_{L}} {\rm{Mpc}}\bigg)^{2}    \bigg (\frac{\Delta V}{\rm{km \ s^{-1}}}\bigg)
\end{equation}\label{eq:HImass}

\noindent
where $z$ is the mean redshift of the stacked sample, $S_{\nu}$ is the mean flux integrated over the $\Delta V$ velocity width of the \HI\ profile in the emitter's frame, and $D_{L}$ is the average luminosity distance. The {\it r}-band is chosen to calculate luminosities for the galaxies, which are then used to evaluate \HI-mass to stellar luminosity ratios.

The advantage of stacking the \HI\ flux in small redshift bins is that one can track any possible evolution of the \HI\ content as function of redshift. However, when exploring the \HI\ properties over the entire redshift range, we stack gas fractions instead of \HI\ flux. This way we compensate for the fact that, for the same amount of \HI, the flux emitted by nearby \HI\ sources is stronger. The extracted spectrum of source $i$ is expressed in terms of \HI\ mass-luminosity fraction: 

\begin{equation}
S_{i} (\nu) \ [Jy] \to S_{i}' (\nu) \  [\rm{Jy} \ \rm{Mpc^{2}} \ \rm{L_{\odot}^{-1}}] =  \frac{2.356 \times 10^{5}} {1+\it{z_{i}}}  \left(\frac{\it{D_{L;i}^{2} S_{i}(\nu)}}  {\it{L_{i}}}\right)
\end{equation}

\noindent
where  $z_{i}$ is the redshift, and $D_{L;i}$ is the luminosity distance, and $L_{i}$ is the $r$-band luminosity of source $i$. After stacking, the mean \HI\ mass-luminosity ratio is integrated over the $\Delta V$ velocity width. 
Before stacking, the average $rms$ noise of the cubes is $\sim$ $0.2$ mJy/beam, which is expected to decrease with the square root of the number of co-added sources.

\section{Characteristics of the galaxies in the selected sample}\label{SampleDescription}

Our main goal is to study/compare the \HI\ properties of various populations of galaxies using large samples with available multiwavelength information. In order to do this, we define several sub-samples, using the collected multiwavelength data.

The color distribution of galaxies is known to consist of two main peaks, i.e. the blue cloud and the red sequence \citep{Strateva2001, Baldry2004}. However, at fixed absolute magnitude the two peaks are not well fit by a double Gaussian due to an excess of galaxies at intermediate colors, also known as the `green valley' \citep{Wyder}. Such intermediate, green colors can be due to a number of different phenomena, e.g. low level (residual) star formation activity \citep{Yi}, dust extinction, older stellar populations \citep{Sarzi}. 

Previous stacking studies, including our LH analysis, were carried out on galaxies separated into blue/red samples, which colors were defined based on ultraviolet-optical NUV $-$ $r$, or optical $g - r$ selections \citep{Fabelloa, Gereb2013}. 
To expand on previous studies, in this paper we consider the green valley as a separate population. Following \cite{Wyder}, in Fig. \ref{fig:CMD} our objects are divided into blue cloud, green valley and red sequence objects based on NUV $-$ $r$ colors. To derive the color distribution of the galaxies, \cite{Wyder} utilizes the fit by \cite{Yi} to the NUV $-$ $r$ colors in function of the $M_r$ absolute magnitude: \mbox{NUV $-$ $r$ = $f$($M_r$) = 1.73 $-$ 0.17$M_r$}. The red sequence is defined as the galaxies with \mbox{NUV $-$ $r$  $>$ $f$($M_r$) $-$ 0.5}, blue galaxies have \mbox{NUV $-$ $r$  $<$ $f$($M_r$) $-$ 2}, whereas green valley galaxies are the excess population between blue and red galaxies with \mbox{$f(M_r) - 2$ $<$ NUV $-$ $r$  $<$ $f$($M_r$) $-$ 0.5} colors. The optical and UV apparent magnitudes are extracted from GALEX and SDSS, and K-corrected following \cite{Chilingarian, Chilingarian2}. The NUV $-$ $r$ colors are corrected for Galactic extinction following \cite{Wyder}.

\begin{figure}
    \begin{center}
      \includegraphics[width=0.43\textwidth]{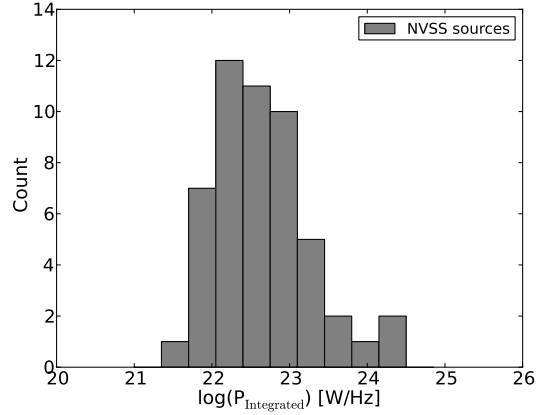}
   \end{center}
\caption{Radio power histogram of NVSS sources in our sample.}
\label{fig:Powerhist}
\end{figure}

From Fig. \ref{fig:CMD} (right) it is clear that in case of the \cite{Gereb2013} selection, our current green valley objects are part of the $g - r > 0.7$ (red) sample, and in the selection of \cite{Fabelloa}, green valley objects belong to the NUV $-$ $r$ < 4.5 (blue) sample.

We also use the Baldwin, Phillips $\&$ Terlevich (BPT) line ratio diagnostic diagram \citep{Baldwin} to separate galaxies with different ionization properties (Fig. \ref{fig:BPT}, left). The selection is done by using line fluxes from SDSS, similar to \cite{Gereb2013}, but with one major difference. Galaxies, which were defined as LINERs in \cite{Gereb2013}, are now separated into composite galaxies and optical AGN samples, using the more stringent demarcation by \cite{Kewley2006} for selecting AGN. Following the BPT classification, our sample includes star-forming (SF) galaxies, composite galaxies where the optical line excitation can come from both star formation and AGN activity, and optical AGN consisting of a mix of Seyfert and Low Ionization Nuclear Emission Region (LINER) galaxies \citep{Kewley2006, Cid2010}. Furthermore, a part (280 sources) of our sample is defined as optically {\sl inactive}, with non-detected, or with maximum two detected lines (among [\ion{N}{II}], H$\alpha$, [\ion{O}{III}], H$\beta$). 
Star-forming galaxies in the BPT diagram in Fig. \ref{fig:BPT} (left) are mainly blue and green, while composite galaxies and optical AGN are green and red. Inactive galaxies are typically red, as in Paper 1.

\begin{table*}[!]
   \begin{center}
   
     \begin{tabular}{c c c c c c c c}

          \hline
               Color       &   Redshift bin   & $M_{\rm HI}$ $(10^{9}$ $\rm M_{\odot})$    & $M{_{\rm HI}}/L_r$ ($\rm M_{\odot}  / \rm L_{\odot} )$ \\
          \hline                            
        
 \multirow{4}{*}{ $g -r < 0.7 $} & $0.02 < z < 0.04 $   &   2.22 $\pm$ 0.05    &   0.45 $\pm$ 0.009  \\
                                                  & $0.04 < z < 0.06 $  &   4.04 $\pm$ 0.07    &   0.48 $\pm$ 0.008   \\                                
                                                  & $0.06 < z < 0.08 $   &  4.71  $\pm$ 0.12  &   0.38  $\pm$ 0.009  \\   
                                                  & $0.08 < z < 0.12 $   &  7.49  $\pm$ 0.24   &   0.39 $\pm$ 0.012  \\  
                                             
   \hline                                                             
                                                       
 \multirow{4}{*}{ $g -r > 0.7 $}   &  $0.02 < z < 0.04 $ & 0.76 $\pm$ 0.04     &  0.06  $\pm$ 0.003 \\
                                                    & $0.04 < z < 0.06 $ &  1.42 $\pm$  0.07    &  0.08 $\pm$ 0.004  \\
                                                    & $0.06 < z < 0.08 $ &  1.81 $\pm$ 0.09    &  0.07 $\pm$ 0.004\\
                                                     & $0.08 < z < 0.12 $ & 3.01 $\pm$ 0.21    &  0.10 $\pm$ 0.007 \\

  \end{tabular}
    \caption{\HI\ mass and mass-luminosity ratio derived for blue ($g -r < 0.7 $) and red ($g -r < 0.7 $) galaxies in the four redshift bins.}\label{table:g-r}
\end{center}
\end{table*}

\begin{figure*}
    \begin{center}
      \includegraphics[width=0.43\textwidth]{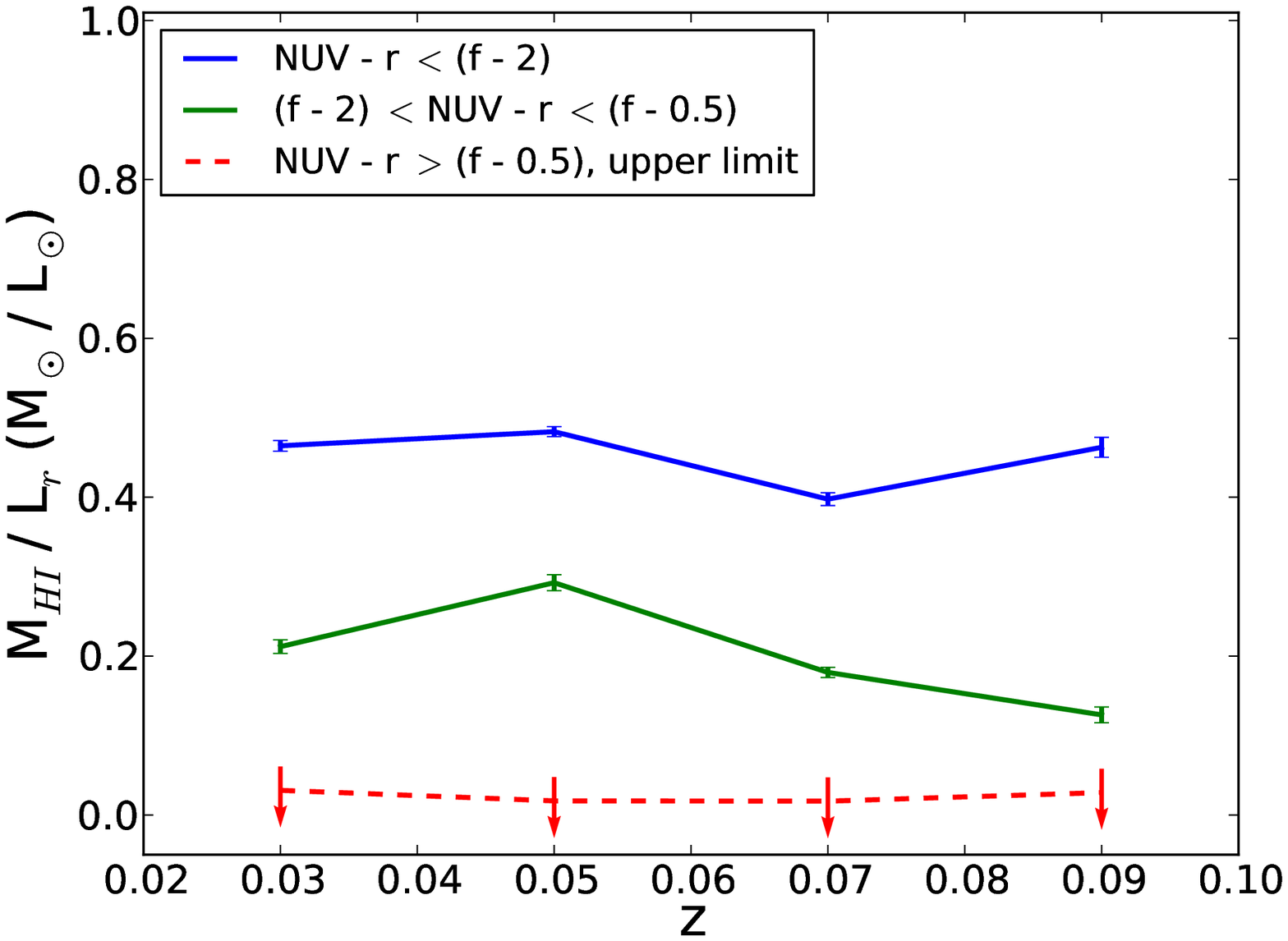}
      \includegraphics[width=0.43\textwidth]{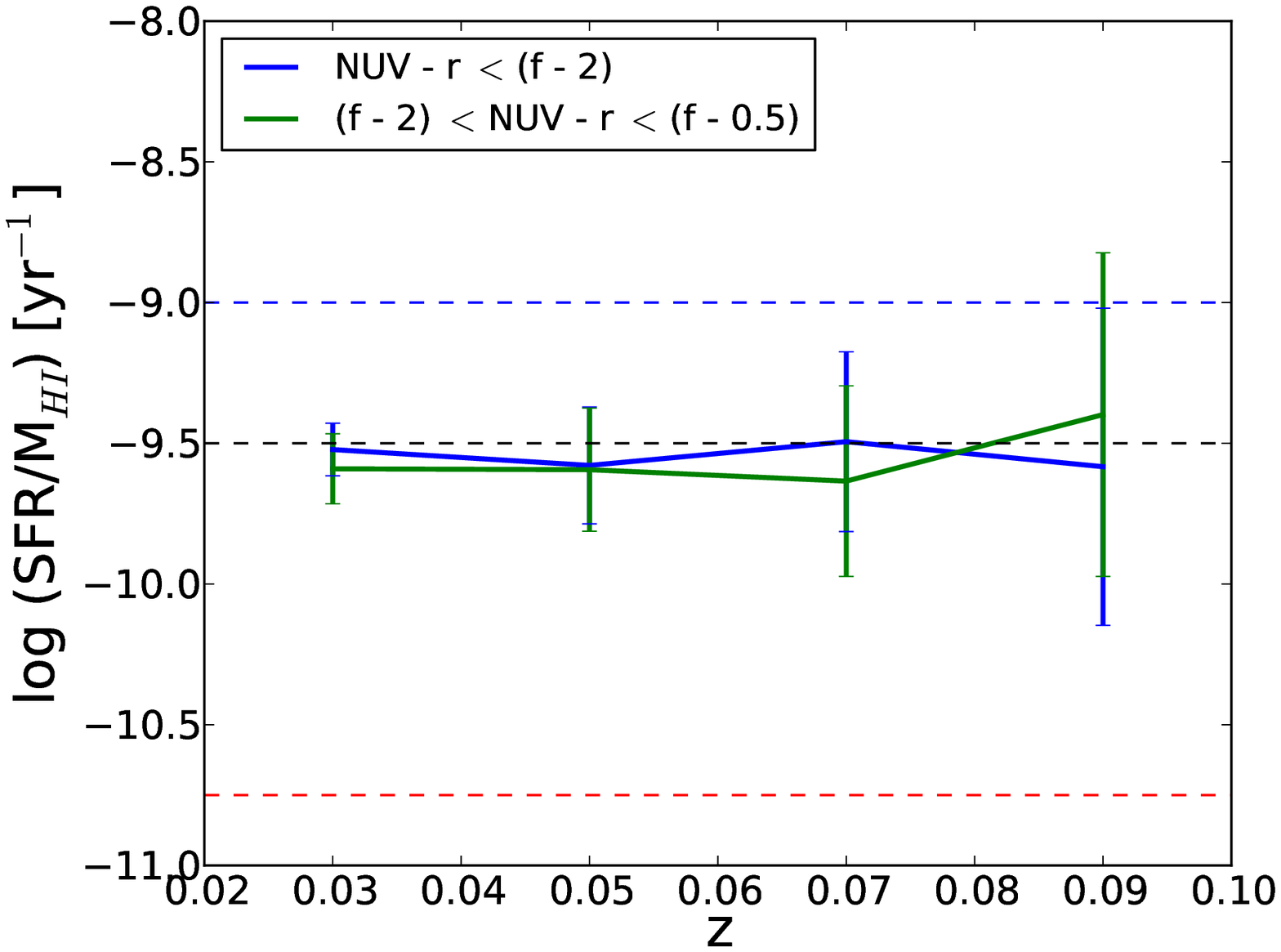}      
      
    \end{center}
\caption{(Left): \HI\ mass-luminosity ratio of galaxies in our current color selection of blue cloud, green valley and red sequence objects. Red arrows and dashed lines indicate the 3-$\sigma$ upper limits.
(Right): The SFE of blue cloud/green valley objects. The blue dashed line indicates the region of high efficiency star formation, whereas the low efficiency region is marked by a red dashed line. The error on the SFR is estimated from the 3-$\sigma$ NUV flux errors.}
 \label{fig:MHI_color}
\end{figure*}

The NVSS radio population is marked by yellow squares in all panels of Fig. \ref{fig:CMD}. Radio sources are among the optically most luminous objects in each population of the color-magnitude diagram in Fig. \ref{fig:CMD} (left). 
From the radio power distribution in Fig. \ref{fig:Powerhist} we expect to have powerful AGN at \mbox{log(P / W Hz$^{-1}$) > 23}, however the distribution shows that low-power radio sources are also present in the sample. In the low radio power regime it becomes more complicated to disentangle the contribution of SF and AGN to the radio continuum emission. 

As previous multiwavelength analysis has shown, IR colors are efficient in separating SF and AGN processes in galaxies (\citealt{Gereb2013} and references therein). With the goal of disentangling SF and AGN activity in our radio sample, we extract 3.4 $\mu$m, 4.6 $\mu$m and 12 $\mu$m magnitudes from WISE to constrain the IR color-color plot, presented in Fig. \ref{fig:BPT} (right). 
The separation at the vertical line (at [4.6$\mu m$] - [12$\mu m$] = 2) in the IR color-color plot (Fig. \ref{fig:BPT}, right) is often used in the literature to disentangle IR early- and late-type galaxies \citep{Wright, Sadler}, or in other words non-star-forming and SF galaxies.
Correspondingly, the radio sources are separated into two distinct populations in Fig. \ref{fig:BPT} (right). A more detailed analysis of this separation is presented in Sec. \ref{Sec:RadioAGN}.

 \begin{figure*}[t!]
    \begin{center}     
          \includegraphics[width=0.43\textwidth]{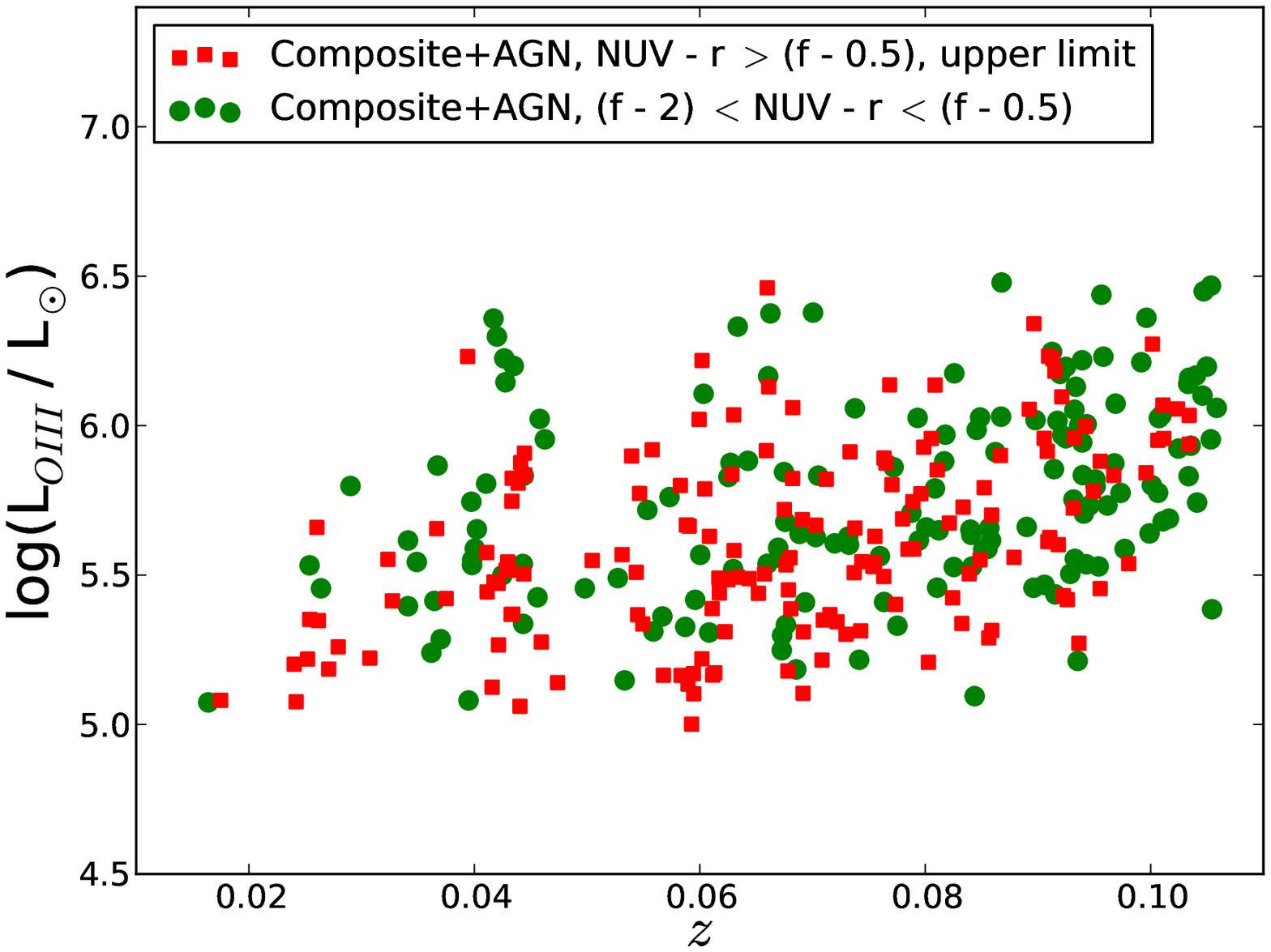}
        \includegraphics[width=0.43\textwidth]{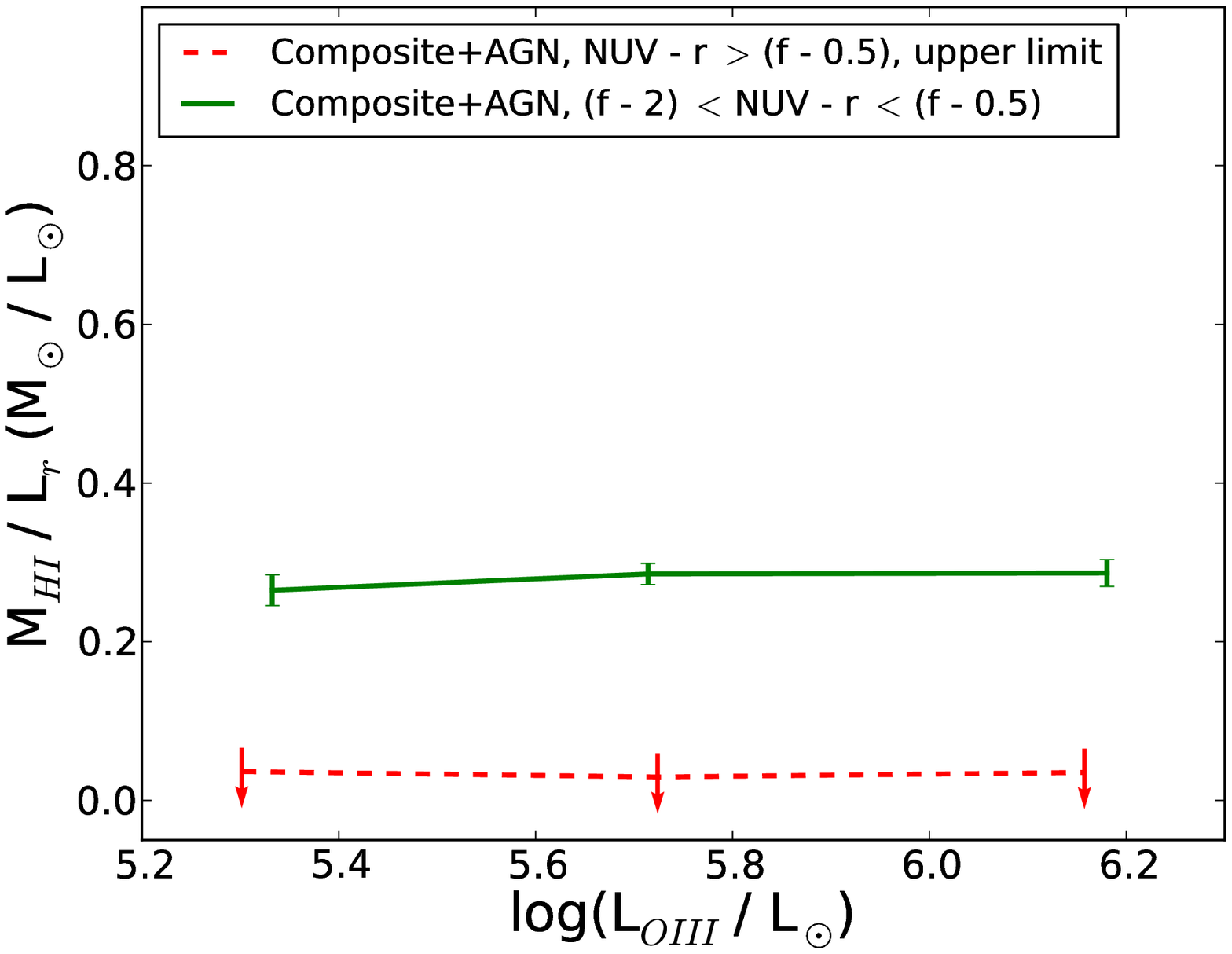}
    \end{center}
\caption{(Left): The [\ion{O}{III}] luminosity distribution of composite galaxies and optical AGN with green and red colors. (Right): \HI\ mass-luminosity ratio of green/red composite galaxies and optical AGN. The 3-$\sigma$ upper limits are marked by a dashed line and arrows. } \label{fig:MHI_LINER_AGN}
\end{figure*}

\section{Results}\label{Results}

\subsection{Stacking in color}\label{Sec:color}\label{Sec:BPT}\label{Sec:RadioAGN}

Before we look at the \HI\ content of blue cloud/green valley/red sequence objects, first we compare our current measurements with the results from the LH study. In the redshift range 0.06 $<$ $z$ $<$ 0.09, in the previous LH field we measured \mbox{M$_{\rm HI}$ $=$ (6.1 $\pm$ 0.4) $\times$ 10$^{9}$ M$_{\odot}$} and M$_{\rm HI}/\rm{L}_r$  $=$ 0.38  $\pm$ 0.02 $\rm M_{\odot}  / \rm L_{\odot} $ in the blue ($g - r$ $<$ 0.7) population, whereas red ($g - r$ $>$ 0.7) galaxies contain lower amounts of gas, with \mbox{M$_{\rm HI}$ $=$ (1.8 $\pm$ 0.2) $\times$ 10$^{9}$ M$_{\odot}$} and M$_{\rm HI}/\rm{L}_r$ $=$ 0.08  $\pm$ 0.01 $\rm M_{\odot}  / \rm L_{\odot} $. 

In Table \ref{table:g-r}, we use $g - r$ optical colors to evaluate the \HI\ mass-luminosity ratio of blue/red galaxies in our current, larger sample. We can average the \HI\ results over the last two redshift bins (0.06 $<$ $z$ $<$ 0.12) to achieve a similar redshift selection as for the LH. In this redshift range, blue galaxies show \mbox{M$_{\rm HI}$ $=$ (6.1 $\pm$ 0.2) $\times$ 10$^{9}$ M$_{\odot}$} and M$_{\rm HI}/\rm{L}_r$  $=$ 0.38  $\pm$ 0.01 $\rm M_{\odot}  / \rm L_{\odot} $, whereas red galaxies have M$_{\rm HI}$ $=$ (2.4 $\pm$ 0.15) $\times$ 10$^{9}$ M$_{\odot}$ and M$_{\rm HI}/\rm{L}_r$  $=$ 0.08  $\pm$ 0.01 $\rm M_{\odot}  / \rm L_{\odot} $. These new measurements confirm the results that we obtained in the LH study.

We now move to the more detailed grouping we can do with the expanded sample. 
\cite{Fabelloa} reported \HI\ detection in red galaxies with NUV $-$ $r$ $>$ 4.5 (similar to our selection of red galaxies), albeit in lower amounts compared to blue galaxies. In Fig. \ref{fig:MHI_color} (left), we present the \HI\ mass-luminosity ratio for blue cloud, green valley and red sequence objects separated based on NUV $-$ $r$ color in our sample. We detect \HI\ in blue and green objects, however unlike in the \cite{Fabelloa} study where more galaxies are available for stacking, red galaxies do not show an \HI\ detection\footnote{We note that in the first redshift bin we find a tentative \HI\ detection in red galaxies at the 3-$\sigma$ level, with M$_{\rm HI}$ $<$ (3.8 $\pm$0.6)  $\times$ 10$^{8}$ M$_{\odot}$ and M$_{\rm HI}/\rm{L}_r$  $<$ 0.03 $\pm$ 0.005 $\rm M_{\odot}  / \rm L_{\odot} $. Higher redshift bins are not detected.} at low 3-$\sigma$ limit of \mbox{M$_{\rm HI}$ $<$ (5 $\pm$ 1.5) $\times$ 10$^{8}$ M$_{\odot}$} and M$_{\rm HI}/\rm{L}_r$  $<$ 0.02 $\pm$ 0.006 $\rm M_{\odot}  / \rm L_{\odot} $ (with values averaged over the four redshift bins).
As we expected based on previous studies, we find that green valley objects are a transition population from \HI\ point of view, showing lower amounts of \HI\ than the blue population, however they are more \HI-rich than the red sample.

Considering that green valley galaxies are expected to have lower amounts of \HI\ than the blue population, it is interesting to explore the timescale over which green valley objects would deplete their gas reservoirs due to star-forming processes. Naively, one could expect that the short transition phase of green valley objects is due to quick consumption of their small gas reservoirs. However, recent results show that in the nearby Universe, the \HI -based star formation efficiency (SFE = SFR/M$_{\rm HI}$) -- or the equivalent inverse, the time scale of neutral gas consumption (t = M$_{\rm HI}$/SFR) -- is independent of other galaxy properties, such as stellar mass, stellar surface density, color, concentration \citep{Schimi, Bigiel}.

Here we test the efficiency of star formation in blue/green galaxies in our sample, baring in mind their \HI\ content. 
The star formation rate (SFR) is derived from the NUV flux, following \cite{Schimi}. The SFR formula accounts for dust attenuation by combining UV-optical colors (NUV $-$ $r$) and the D$_n$(4000) index of galaxies. The latter index is an indicator of the presence of young stellar populations. 
In each redshift bin, the \HI-based star formation efficiency is defined as the average SFR over stacked \HI\ mass, i.e.  \mbox{<$\rm SFR$>/<$M_{\rm{\HI}}$>}. 

In Fig. \ref{fig:MHI_color} (right), green valley objects and blue galaxies reveal similar, efficient star formation, with little variations around \mbox{SFE = 10$^{-9.5}$ yr$^{-1}$}, corresponding to a gas consumption time scale of t $\sim$ 3$\times$10$^{9}$ yr. Red galaxies (not shown) lack any sign of \HI\ gas and star formation activity. Our estimates for the SFR and SFE agree well with the results of previous studies \citep{Schimi}. The similar SFE detected in blue/green galaxies suggests that the \HI-based efficiency of star formation is independent of color. 

Our current study spans quite a large redshift range with respect to the majority of previous \HI\ studies. Stacking our galaxies in redshift bins allows us to trace the global \HI\ properties as function of redshift. We note that the global \HI\ content (mass-luminosity ratio) does not change significantly up to $z \sim 0.1$. Even though the sampled redshift range here is too small to relate our results to the cosmological evolution of \HI, the constancy of the global \HI\ content as function of redshift up to $z \sim 0.1$ is consistent with what previous studies have found \citep{Freudling2011, Delhaize}.
Similarly to the behaviour of the global \HI\ content, the SFE shows only little variations with redshift up to $z \sim 0.1$. Over the probed redshift range, the gas consumption timescale of $\sim$1 - 3 Gyr is comparable to the lookback time of $\sim$1.5 Gyr. Given that the timescales are very similar, gas consumption by star-forming processes should have a detectable effect on the global \HI\ content. However, the constancy of the \HI\ mass-luminosity ratio and SFE as function of redshift is suggestive that the gas in galaxies is replenished on a timescale of a few Gyr-s. Similar conclusions for the need of gas replenishment have been reached for young galaxies between $z = 2$ and $z = 1$ \citep{Tacconi2010}.

\subsection{The \HI\ properties of composite galaxies and optical AGN}\label{Sec:BPT}\label{Sec:RadioAGN}

Here we want to test the effects that are responsible for the decreased \HI\ content in optically selected AGN.                         
Is the lower \HI\ content in composite galaxies/AGN the result of galaxies being redder (with older stellar populations, dustier), or it is related to ionization/AGN feedback properties?

In the following analysis we select a combined sample of composite galaxies and optical AGN considering all galaxies above the dashed line in the BPT diagram in Fig. \ref{fig:BPT} (left), and we use NUV $-$ $r$ colors to test the color dependence of \HI\ in the selected populations.
The strength of the [\ion{O}{III}] emission line is a fairly reliable tracer of black hole accretion \citep{Kauffmann}. To achieve a uniform emission line selection, we consider galaxies with [\ion{O}{III}] luminosities between \mbox{5 $<$ log(L$_{ [\ion{O}{III}] } / L_{\odot})  $  $<$ 6.5}. Following this selection, less than 10$\%$ of the AGN are excluded from the sample. We examine the color of the selected objects, and we find that they are mostly green and red. In Fig. \ref{fig:MHI_LINER_AGN} (left) we show the [\ion{O}{III}] luminosity distribution of the selected green/red AGN as function of redshift. 

To test the \HI\ properties of our AGN as function of color and nuclear activity, we stack the gas fractions of red and green AGN separately as function of their [\ion{O}{III}] luminosity. The stacking is performed in three luminosity bins with steps of \mbox{$\Delta$log$(L_{ [\ion{O}{III}] } / L_{\odot}) $ = 0.5}. The results of this stacking experiment are shown in Fig. \ref{fig:MHI_LINER_AGN} (right). 
The \HI\ is concentrated in green objects among composite galaxies/AGN, whereas red galaxies are not detected at the 3-$\sigma$ level of M$_{\rm HI}/\rm{L}_r$ $<$ 0.03 $\pm$ 0.01 $\rm M_{\odot}  / \rm L_{\odot} $. This suggests that even those galaxies in which the presence of an AGN is expected to be more likely (composite galaxies/AGN) do show \HI\ detections, however this depends on their color. Furthermore, the \HI\ remains relatively constant as function of [\ion{O}{III}] luminosity. This is a strong indication that the \HI\ content is well correlated with the NUV $-$ $r$ color and the SF history of the galaxies, while the effect of optical AGN feedback on the diffuse atomic gas is less significant. Similar conclusion has been reached in previous works by \cite{Fabellob} and \cite{Ho2008}.

\subsection{AGN and SF properties of the radio population}\label{Sec:RadioAGN}

The two main processes that produce radio emission in galaxies are AGN and SF. In the nearby Universe, the largest black holes are hosted by massive galaxies, which generally have older stellar populations and are poor in gas. Optically these are not very powerful AGN, which is also reflected by the fact that about 40$\%$ of our radio AGN sample is optically inactive with [\ion{O}{III}] emission lines below the 3-$\sigma$ significance level. However, radio selection is sensitive to detecting this type of AGN. Because radio-loud AGN can not be identified based on emission lines, we need to look for other multiwavelength diagnostics.

As we show in Sec \ref{Sec:BPT}, radio sources are divided into two well separated groups in the IR color-color diagram, e.g. IR late-type galaxies (on the right) and IR early-type galaxies (on the left). In Fig. \ref{fig:BPT} (right), the IR late-type region is populated by star-forming and composite galaxies, whereas the IR early-type region is dominated by optical AGN and non-star-forming (optically inactive) galaxies. Radio AGN are most likely to be found in the latter sample.

\begin{figure}
   \begin{center}     
        \includegraphics[width=0.43\textwidth]{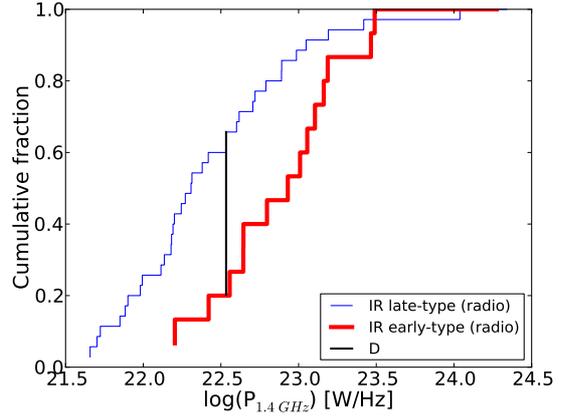}
    \end{center}
\caption{Cumulative fraction of the radio power distribution in IR late-type and IR early-type galaxies. $D$ is the maximum distance between the cumulative fraction of the two distributions. We measure D = 0.504 for 35 IR late-type, and 15 IR early-type galaxies. } \label{fig:KS}
\end{figure}

As radio AGN are expected to be more powerful than SF galaxies, in Fig. \ref{fig:KS} we compare the radio power distribution of IR late-type and IR early-type radio sources. The two distributions have different shapes, with a wide, D = 0.504 maximum distance between the two cumulative distribution functions. According to the Kolmogorov-Smirnov test, the probability that the two distributions are different is 99\%, implying that statistically IR late-type and IR early-type galaxies have a different radio power distribution. The mean radio power of IR late-type galaxies is \mbox{log(P / W Hz$^{-1}$) = 22.5}, whereas as expected, IR early-type galaxies are typically more powerful, with a mean \mbox{log(P / W Hz$^{-1}$) = 23}.   

\begin{figure}
   \begin{center}     
        \includegraphics[width=0.43\textwidth]{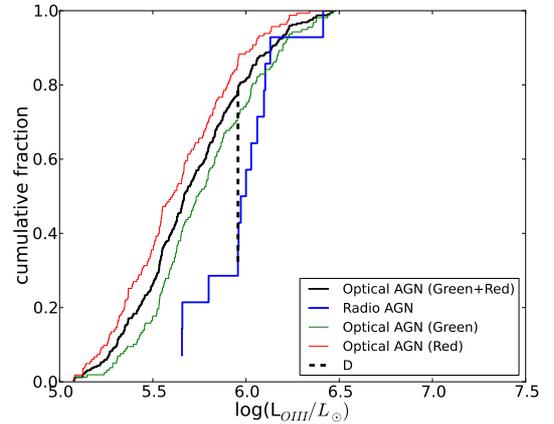}
    \end{center}
\caption{Cumulative fraction of the [\ion{O}{III}] luminosity distribution of all optically selected AGN (red and green composite+AGN) in black, and radio-selected AGN in blue. Red and green optical AGN are also marked separately. We measure the maximum distance between the distribution of 321 optical AGN and 14 radio sources as D = 0.564. }\label{fig:KS_LOIII}
\end{figure}

We want to test whether the radio AGN candidates and optically selected AGN population show similar properties in terms of  [\ion{O}{III}] luminosity and black hole accretion, or the two AGN phenomena are not related. In Fig. \ref{fig:KS_LOIII} we compare the [\ion{O}{III}] luminosity distribution of the two samples using the Kolmogorov-Smirnov test. For the AGN which are optically inactive, the 3-$\sigma$ upper limit is considered when calculating the cumulative distribution. The probability that the two samples are drawn from the same distributions is 0.01$\%$. In agreement with \cite{Best2005}, this result implies that radio AGN luminosity and optical AGN emission lines are independent from each other.

In Paper 1. we carried out \HI\ stacking on radio sources separated into SF and AGN samples in a similar way, by IR colors. We found that radio-emitting SF galaxies contain \HI, whereas we did not detect \HI\ in low-luminosity radio AGN at the \mbox{M$_{\rm HI}$ $<$ (1.2 $\pm$ 0.40) $\times$ 10$^{9}$ M$_{\odot}$} upper limit. Now we do the same analysis with radio-selected NVSS objects.

Stacking results in an \HI\ detection in IR late-type galaxies, showing M$_{\rm HI}$ $=$ (3.04 $\pm$ 0.37) $\times$ 10$^{9}$ M$_{\odot}$ and \mbox{M$_{\rm HI}/\rm{L}_r$  $=$ 0.11 $\pm$ 0.01 $\rm M_{\odot}  / \rm L_{\odot}$} (the values are averaged over the entire redshift range). The \HI\ mass-luminosity ratios of these, likely star forming galaxies is lower with respect to the entire blue SF population in Fig. \ref{fig:MHI_color}. The low mass-luminosity ratio is partly the result of the high optical luminosity of radio-detected objects, which are among the brightest sources in the color-magnitude diagram in Fig. \ref{fig:CMD} (left plot). It is interesting to note that the IR late-type radio population reveals a high SFE of $\sim$10$^{-9}$ yr$^{-1}$, corresponding to gas consumption time scales of \mbox{t = 10$^{9}$ yr}. This SFE is more enhanced than what we detect for the entire SF population. Likely, the radio emission in these sources is the result of their enhanced star formation activity.

In the IR early-type region stacking reveals an \HI\ non-detection with a mass and mass-luminosity upper limit of \mbox{M$_{\rm HI}$ $<$ (1.95 $\pm$ 0.6) $\times$ 10$^{9}$ M$_{\odot}$} and M$_{\rm HI}/\rm{L}_r$  $<$ 0.03 $\pm$ 0.01 $\rm M_{\odot}  / \rm L_{\odot} $.  Along with the high radio power and lack of star formation (based on their red colors and location in the IR color-color diagram), the \HI\ non-detection supports that radio emission in the IR early-type region is due to radio AGN.

\section{Discussion and summary}\label{Discussion}

\begin{itemize} 

\item[$\bullet$] Our stacking results show that galaxies in the green valley are detected with lower amounts of \HI\ than blue galaxies, but unlike red galaxies, they are not completely depleted of cold (\HI) gas. This result can be expected based on earlier studies of \HI\ scaling relations  \citep{Catinella2010, Catinella2012}. This result suggests that SF goes hand in hand with the \HI\ properties, and in galaxies where cold \HI\ gas is present, conditions are favourable for (residual) SF to be seen. 

\item[$\bullet$] In Fig. \ref{fig:MHI_color}, the \HI\ mass-luminosity ratio does not change significantly as function of redshift, suggesting that the \HI\ content remains relatively constant up to $z \sim 0.1$. Over our relatively small redshift range, this result agrees with other studies of the global \HI\ content.  Furthermore, the SFE displays a similar behaviour, remaining relatively constant in the covered redshift range. It seems that the \HI-based SFE does not depend on color, meaning that green galaxies consume their \HI\ on the same timescale as blue galaxies. The fact that the \HI-based SFE does not seem to depend on galaxy properties has been interpreted as an indications that the \HI\ content and SF are regulated by the same process, e.g. feedback effects, galaxy environment \citep{Schimi}. On the other hand, it has been shown that the H$_2$-based SFE does depend on several galaxy properties.  Systematic studies of the atomic and molecular gas content of galaxies show that there is a significant scatter between the amount of \HI\ and H$_2$, which explains while there are clear trends between M$_{\rm H_{2}} $ and SFR, but far weaker ones between M$_{\rm HI}$ and SFR \citep{Saintonge2011, Saintonge2012}. As such, these studies argue that the lack of dependencies of the \HI-based SFE may rather reflect that there is a wide range in the structural properties of the \HI-rich outer disks of galaxies. Perhaps, the \HI\ and SF properties of galaxies depend both on environmental/feedback effects and structural properties of the gaseous disks. Future, high-resolution \HI\ surveys will make it possible to study the relation of \HI\ to SF properties as function of structural/environmental/feedback effects. Future surveys will also allow for tracing the evolution of the global \HI\ content as function of redshift, by extending \HI\ studies to cosmologically significant distances. 

\item[$\bullet$] We detect \HI\ gas in green optical AGN, suggesting that even galaxies with higher ionization properties (composite galaxies and optical AGN) do contain neutral gas. However, red AGN in our sample are not detected in \HI. The \HI\ mass-luminosity ratio remains relatively constant as function of [\ion{O}{III}] luminosity, suggesting that optical AGN are not the (main) reason for depleting gas reservoirs. In agreement with previous studies, our results show that the presence of \HI\ is better correlated with NUV $-$ $r$ color rather than with ionization properties.

\item[$\bullet$] In radio sources located in the IR late-type region, small amounts of gas are associated with very efficient star formation. The lack of \HI\ and the high level of residual star formation suggest that these galaxies recently went through an intense star-formation period, and this led to a significant depletion of \HI\ in these galaxies. 

\item[$\bullet$] We do not detect any \HI\ gas in radio sources located in the IR early-type region ([4.6 $\mu m$] - [12 $\mu m$] $<$ 2) in the WISE color-color plot. The lack of \HI\ gas along with the non-star-forming properties and high average radio power of \mbox{log(P / W Hz$^{-1}$) = 23} suggest that the radio emission in this population can not originate from star formation. Therefore, radio AGN are likely to be responsible for the radio continuum emission in IR early-type galaxies. 

\item[$\bullet$] Radio-selected AGN have a different [\ion{O}{III}] luminosity distribution than optically defined AGN. Following \cite{Best2005}, optical and radio AGN represent two different phenomena, triggered in a different way. Unlike optically selected AGN, the radio AGN population is expected to be able to deplete/heat cold gas reservoirs (see Introduction). On the red sequence, both the optical and radio AGN populations have an \HI\ non-detection, and our current sample is too small to test any differences in the \HI\ properties of the two AGN populations at low detection limit. However, it will be interesting to explore this at lower limit with future surveys.
In our current selection, we reach a detection limit of \mbox{M$_{\rm HI}$ $<$ (5 $\pm$ 1.5) $\times$ 10$^{8}$ M$_{\odot}$} and \mbox{M$_{\rm HI}/\rm{L}_r$  $<$ 0.02 $\pm$ 0.006 $\rm M_{\odot}  / \rm L_{\odot} $} (averaged over four redshift bins) in the red sample. Even though this is a relatively low detection limit, lower \HI\ masses have been detected before by direct observations of the \sauron\ and \atlas\ samples, and by earlier stacking studies of ALFALFA galaxies \citep{Fabelloa, Fabellob}. 
Stacking is a promising technique to lower the detection limit and study the relatively unexplored $<$ 10$^{7}$ M$_{\odot}$ \HI\ mass regime of galaxies using large samples of galaxies. This will be made possible by future \HI\ surveys with the next generation of radio telescopes, e.g. Apertif \citep{Oosterloo2010b}), the Australian Square Kilometre Array Pathfinder (ASKAP, \citealt{DeBoer2009}), MeerKat \citep{Booth2009}, and the Karl G. Jansky Very Large Array (JVLA, e.g. CHILES survey,  \citealt{CHILES2013}).

\end{itemize}

\section{Acknowledgements} 
We thank the referee for the useful and detailed comments that helped us to improve the manuscript.\\
The WSRT is operated by the ASTRON (Netherlands Foundation for Research in Astronomy) with support from the Netherlands Foundation
for Scientific Research (NWO).\\
RM gratefully acknowledge support from the European Research Council under the European Union's Seventh Framework Programme (FP/2007-2013) /ERC Advanced Grant RADIOLIFE-320745.\\
This research made use of the ``K-corrections calculator'' service available at http://kcor.sai.msu.ru/


\begin{thebibliography}{Vermeulen et al.(2008)}

\bibitem[Baldry et al.(2004)]{Baldry2004} Baldry, I.~K., Glazebrook, K., Brinkmann, J., et al.\ 2004, \apj, 600, 681 

\bibitem[\protect\citeauthoryear{Baldwin, Phillips, \& Terlevich}{1981}]{Baldwin} Baldwin J.~A., Phillips M.~M., Terlevich R., 1981, PASP, 93, 5

\bibitem[Best et al.(2005)]{Best2005} Best, P.~N., Kauffmann, G., Heckman, T.~M., et al.\ 2005, \mnras, 362, 25 


\bibitem[Best \& Heckman(2012)]{Best2012} Best, P.~N., \& Heckman, T.~M.\ 2012, \mnras, 421, 1569 

\bibitem[Bigiel et al.(2011)]{Bigiel} Bigiel, F., Leroy, A.~K., Walter, F., et al.\ 2011, \apjl, 730, L13 

\bibitem[DeBoer et al.(2009)]{DeBoer2009} DeBoer, D.~R., Gough, R.~G., Bunton, J.~D., et al.\ 2009, IEEE Proceedings, 97, 1507 

\bibitem[Debuhr et al.(2012)]{Debuhr2012} Debuhr, J., Quataert, E., \& Ma, C.-P.\ 2012, \mnras, 420, 2221 

\bibitem[Booth et al.(2009)]{Booth2009} Booth, R.~S., de Blok, W.~J.~G., Jonas, J.~L., \& Fanaroff, B.\ 2009, arXiv:0910.2935 

\bibitem[Booth \& Schaye(2009)]{Booth2009Feedb} Booth, C.~M., \& Schaye, J.\ 2009, \mnras, 398, 53 

\bibitem[Catinella et al.(2010)]{Catinella2010} Catinella, B., Schiminovich, D., Kauffmann, G., et al.\ 2010, \mnras, 403, 683 

\bibitem[Catinella et al.(2012)]{Catinella2012} Catinella, B., Schiminovich, D., Kauffmann, G., et al.\ 2012, \aap, 544, AA65 

\bibitem[Catinella \& Cortese(2015)]{Catinella2015} Catinella, B., \& Cortese, L.\ 2015, \mnras, 446, 3526 

\bibitem[Chilingarian et al.(2010)]{Chilingarian} Chilingarian, I.~V., Melchior, A.-L., \& Zolotukhin, I.~Y.\ 2010, \mnras, 405, 1409 

\bibitem[Chilingarian \& Zolotukhin(2012)]{Chilingarian2} Chilingarian, I.~V., \& Zolotukhin, I.~Y.\ 2012, \mnras, 419, 1727 

\bibitem[Cid Fernandes et al.(2010)]{Cid2010} Cid Fernandes, R., Stasi{\'n}ska, G., Schlickmann, M.~S., et al.\ 2010, \mnras, 403, 1036 


\bibitem[Condon et al.(1998)]{Condon1998} Condon, J.~J., Cotton, W.~D., Greisen, E.~W., et al.\ 1998, \aj, 115, 1693 

\bibitem[Cortese et al.(2011)]{Cortese2011} Cortese, L., Catinella, B., Boissier, S., Boselli, A., \& Heinis, S.\ 2011, \mnras, 415, 1797 


\bibitem[Croton(2006)]{Croton2006} Croton, D.~J.\ 2006, \mnras, 369, 1808 


\bibitem[\protect\citeauthoryear{Delhaize et al.}{2013}]{Delhaize} Delhaize J., Meyer M.~J., Staveley-Smith L., Boyle B.~J., 2013, MNRAS, 433, 1398 

\bibitem[De Lucia \& Blaizot(2007)]{deLucia2007} De Lucia, G., \& Blaizot, J.\ 2007, \mnras, 375, 2 

\bibitem[Di Matteo et al.(2005)]{diMatteo2005} Di Matteo, T., Springel, V., \& Hernquist, L.\ 2005, \nat, 433, 604 


\bibitem[\protect\citeauthoryear{Fabello et al.}{2011a}]{Fabelloa} Fabello S., Catinella B., Giovanelli R. et al., 2011, MNRAS, 411, 993 

\bibitem[\protect\citeauthoryear{Fabello et al.}{2011b}]{Fabellob} Fabello S., Kauffmann G., Catinella B. et al., 2011, MNRAS, 416, 1739 

\bibitem[Fernandez et al.(2015)]{CHILES2013} Fernandez, X., van Gorkom, J.~H., Momjian, E., \& Chiles Team 2015, American Astronomical Society Meeting Abstracts, 225, \#427.03 

\bibitem[Freudling et al.(2011)]{Freudling2011} Freudling, W., Staveley-Smith, L., Catinella, B., et al.\ 2011, \apj, 727, 40 

\bibitem[Ger{\'e}b et al.(2013)]{Gereb2013} Ger{\'e}b, K., Morganti, R., Oosterloo, T.~A., Guglielmino, G., \& Prandoni, I.\ 2013, \aap, 558, A54 

\bibitem[Giovanelli et al.(2005)]{Giovanelli2005} Giovanelli, R., Haynes, M.~P., Kent, B.~R., et al.\ 2005, \aj, 130, 2598 

\bibitem[Haynes et al.(2011)]{Haynes2011} Haynes, M.~P., Giovanelli, R., Martin, A.~M., et al.\ 2011, \aj, 142, 170 

\bibitem[Ho et al.(2008)]{Ho2008} Ho, L.~C., Darling, J., \& Greene, J.~E.\ 2008, \apj, 681, 128 

\bibitem[Hopkins et al.(2006)]{Hopkins2006Feedb} Hopkins, P.~F., Hernquist, L., Cox, T.~J., et al.\ 2006, \apjs, 163, 1 

\bibitem[\protect\citeauthoryear{van der Hulst, van Albada, \& Sancisi}{2001}]{Hulst} van der Hulst J.~M., van Albada T.~S., Sancisi R., 2001, ASPC, 240, 451 

\bibitem[\protect\citeauthoryear{Kauffmann et al.}{2003}]{Kauffmann} Kauffmann G., et al., 2003, MNRAS, 346, 1055 

\bibitem[\protect\citeauthoryear{Kennicutt}{1998}]{Kennicutt1998} Kennicutt R.~C., Jr., 1998, ApJ, 498, 541 

\bibitem[\protect\citeauthoryear{Kewley et al.}{2001}]{Kewley} Kewley L.~J., Heisler C.~A., Dopita M.~A., Lumsden S., 2001, ApJS, 132, 37 

\bibitem[\protect\citeauthoryear{Kewley et al.}{2006}]{Kewley2006} Kewley L.~J., Groves B., Kauffmann G., Heckman T., 2006, MNRAS, 372, 961 

\bibitem[Lah et al.(2007)]{Lah2007} Lah, P., Chengalur, J.~N., Briggs, F.~H., et al.\ 2007, \mnras, 376, 1357 

\bibitem[Lah et al.(2009)]{Lah2009} Lah, P., Pracy, M.~B., Chengalur, J.~N., et al.\ 2009, \mnras, 399, 1447 

\bibitem[\protect\citeauthoryear{Meyer et al.}{2004}]{Meyer2004} Meyer M.~J., et al., 2004, MNRAS, 350, 1195 

\bibitem[Martin et al.(2007)]{Martin2007} Martin, D.~C., Small, T., Schiminovich, D., et al.\ 2007, \apjs, 173, 415 

\bibitem[\protect\citeauthoryear{Morganti et al.}{2006}]{Morganti2006} Morganti R., et al., 2006, MNRAS, 371, 157 

\bibitem[\protect\citeauthoryear{Oosterloo et al.}{2010}]{Oosterloo2010} Oosterloo T., et al., 2010, MNRAS, 409, 500 

\bibitem[Oosterloo et al.(2010b)]{Oosterloo2010b} Oosterloo, T., Verheijen, M., \& van Cappellen, W.\ 2010, ISKAF2010 Science Meeting

\bibitem[\protect\citeauthoryear{Sadler et al.}{2013}]{Sadler} Sadler E.~M., Ekers R.~D., Mahony E., Mauch T., Murphy T., 2013, arXiv, arXiv:1304.0268 

\bibitem[Saintonge et al.(2011)]{Saintonge2011} Saintonge, A., Kauffmann, G., Wang, J., et al.\ 2011, \mnras, 415, 61 

\bibitem[Saintonge et al.(2012)]{Saintonge2012} Saintonge, A., Tacconi, L.~J., Fabello, S., et al.\ 2012, \apj, 758, 73 


\bibitem[Salim et al.(2007)]{Salim2007} Salim, S., Rich, R.~M., Charlot, S., et al.\ 2007, \apjs, 173, 267 

\bibitem[\protect\citeauthoryear{Sarzi et al.}{2010}]{Sarzi} Sarzi M., et al., 2010, MNRAS, 402, 2187 

\bibitem[\protect\citeauthoryear{Sault, Teuben, \& Wright}{1995}]{Sault} Sault R.~J., Teuben P.~J., Wright M.~C.~H., 1995, ASPC, 77, 433 

\bibitem[Schiminovich et al.(2007)]{Schimi2007} Schiminovich, D., Wyder, T.~K., Martin, D.~C., et al.\ 2007, \apjs, 173, 315

\bibitem[\protect\citeauthoryear{Schiminovich et al.}{2010}]{Schimi} Schiminovich D., et al., 2010, MNRAS, 408, 919 

\bibitem[\protect\citeauthoryear{Serra et al.}{2012}]{Serra} Serra P., et al., 2012, MNRAS, 422, 1835 

\bibitem[Somerville et al.(2008)]{Somerville2008} Somerville, R.~S., Hopkins, P.~F., Cox, T.~J., Robertson, B.~E., 
\& Hernquist, L.\ 2008, \mnras, 391, 481 

\bibitem[Strateva et al.(2001)]{Strateva2001} Strateva, I., Ivezi{\'c}, {\v Z}., Knapp, G.~R., et al.\ 2001, \aj, 122, 1861 

\bibitem[Tacconi et al.(2010)]{Tacconi2010} Tacconi, L.~J., Genzel, R., Neri, R., et al.\ 2010, \nat, 463, 781 


\bibitem[\protect\citeauthoryear{Verheijen et al.}{2007}]{Verheijen} Verheijen M., van Gorkom J.~H., Szomoru A., Dwarakanath K.~S., Poggianti B.~M., Schiminovich D., 2007, ApJ, 668, L9 

 \bibitem[Wagner et al.(2012)]{Wagner} Wagner, A.~Y., Bicknell, G.~V., \& Umemura, M.\ 2012, \apj, 757, 136 
 
\bibitem[\protect\citeauthoryear{Walter et al.}{2008}]{THINGS}  Walter F., Brinks E., de Blok W.~J.~G., Bigiel F., Kennicutt R.~C., Jr., Thornley M.~D., Leroy A., 2008, AJ, 136, 2563 

\bibitem[\protect\citeauthoryear{Wright et al.}{2010}]{Wright} Wright E.~L., et al., 2010, AJ, 140, 1868 

\bibitem[\protect\citeauthoryear{Wyder et al.}{2007}]{Wyder} Wyder T.~K., et al., 2007, ApJS, 173, 293 

\bibitem[Yi et al.(2005)]{Yi} Yi, S.~K., Yoon, S.-J., Kaviraj, S., et al.\ 2005, \apjl, 619, L111 

\bibitem[\protect\citeauthoryear{York et al.}{2000}]{York} York D.~G., et al., 2000, AJ, 120, 1579 

\bibitem[Zhang et al.(2009)]{Zhang2009} Zhang, W., Li, C., Kauffmann, G., et al.\ 2009, \mnras, 397, 1243 

\bibitem[\protect\citeauthoryear{Zwaan et al.}{2005}]{Zwaan2005} Zwaan M.~A., Meyer M.~J., Staveley-Smith L., Webster R.~L., 2005, MNRAS, 359, L30 


\end{thebibliography}
\end{document}